\documentclass[english,12pt]{cern2}

\usepackage{amssymb}
\usepackage{graphicx}
\usepackage{babel}
\usepackage{multicol}
\usepackage{setspace}

\begin{document}

\title{Surface Analysis of OFE-Copper X-Band Accelerating Structures and
Possible Correlation to RF Breakdown Events}
\author{S.E. Harvey\thanks{Corresponding Author: stef@slac.stanford.edu}, F. Le Pimpec,
R.E. Kirby, F.Marcelja, K. Adamson, E.L. Garwin \\
SLAC, 2575 Sand Hill Rd, Menlo Park, CA, 94025}

\maketitle

\thispagestyle{myheadings}
\markright{AS-WeP21}

\begin{abstract}
X-band accelerator structures meeting the Next Linear Collider
(NLC) design requirements have been found to suffer vacuum surface
damage caused by radio frequency (RF) breakdown, when processed to
high electric-field gradients. Improved understanding of these
breakdown events is desirable for the development of structure
designs, fabrication procedures, and processing techniques that
minimize structure damage. RF reflected wave analysis and acoustic
sensor pickup have provided breakdowns localization in RF
structures \cite{Nelson:PAC03} \cite{adolphsen:pac03}. Particle
contaminations found following clean autopsy of four RF-processed
travelling wave structures, have been catalogued and analyzed.
Their influence on RF breakdown, as well as that of several other
material-based properties, will be discussed.
\end{abstract}

\doublespacing

\section{Introduction}

RF breakdown rates in Next Linear Collider (NLC) X-band
(11.424~GHz) travelling wave structures currently exceed the
design limits (i.e. one breakdown per 10~hours at an electric
field of 65MV/m with a 400ns pulse length). In addition to
excessive breakdown rates, a high degree of variability in
performance among identically manufactured and processed RF
structures has been observed.

In the late 1990s NLC structure prototype testing began with the
availability of more powerful RF sources; initial tests revealed
catastrophic RF breakdown surface damage and phase shift
(detuning) beginning at an average gradient of 50MV/m
\cite{adolphsen:Linac2000} \cite{adolphsen:pac01}. As RF breakdown
performance has become one of several critical factors defining
the limit of the operating gradient of an accelerating structure,
greater efforts have been made in characterizing, understanding,
and reducing RF breakdowns. A key question which needs to be
addressed in depth is the limiting factors on the use of copper as
an accelerator material, as accelerating gradients move toward
100MV/m.

\begin{figure}[tbph]
\begin{center}
\includegraphics[width=0.8\textwidth,clip=]{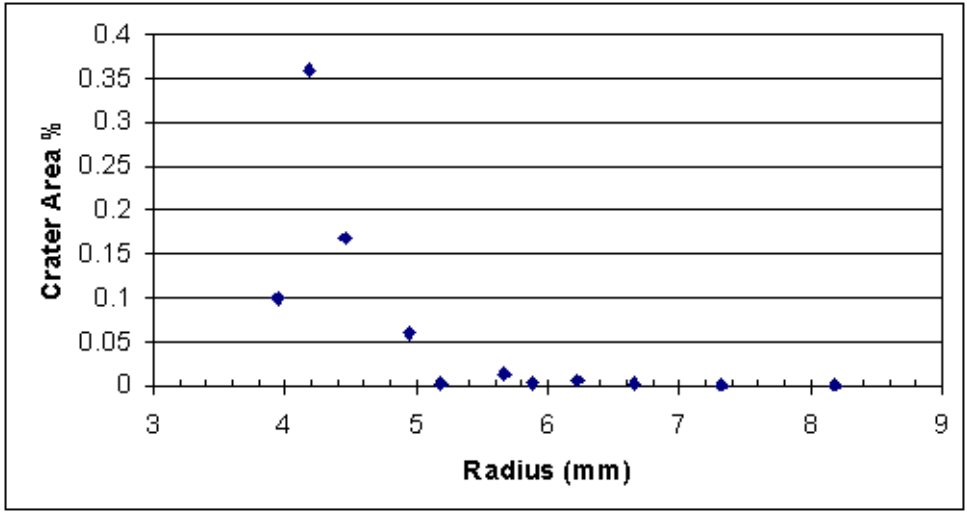}
\end{center}
\caption{Average percentage area damaged by craters versus radial
distance from center of aperture (30$^{th}$Cell T53VG3RA)}
\label{figcraterdensity}
\end{figure}

This paper explores the possible correlation between the location
and frequency of occurrence of breakdown events
(Fig.\ref{figsharpsmooth}(a and b) and Fig.\ref{figcraterdensity})
and their connection to copper surface morphology or contamination
in travelling wave structures. The goal of this ongoing work is to
identify the cause of the breakdown events and to diminish their
occurrence, hence reducing operational variation among structures.

The universally accepted starting point for the initiation of RF
or DC breakdown is field emission. RF breakdown events take place
when the local field emission current density at a given site
causes ohmic-heating sufficient to locally release gas, leading to
ionization and ion back-bombardment that leads to further heating,
plasma formation and, occasionally, catastrophic breakdown
\cite{llaurent} \cite{Wilson:pac03}.

\section{Accelerator Structures}

\begin{figure}[tbph]
\begin{center}
\includegraphics*[width=0.8\textwidth,clip=]{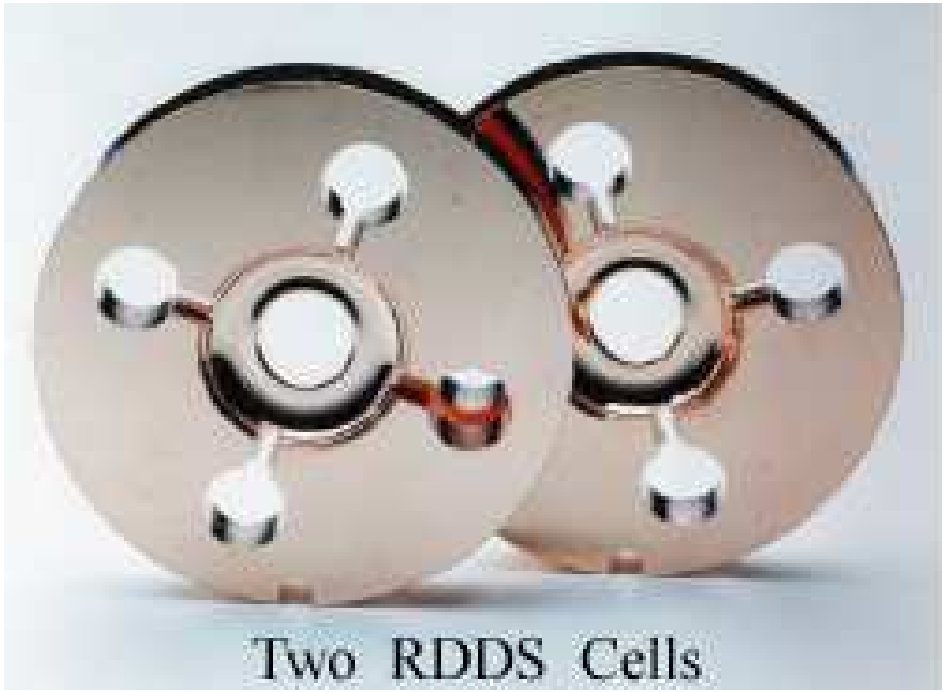}
\end{center}
\caption{Damped Round and Detuned Structure Cells}
\label{figRDDScells}
\end{figure}

The accelerator structure is manufactured from 99.99{\%} pure OFE
copper that has been evaluated as Class 1 material per ASTM-B66.
To form a cell, Fig.\ref{figRDDScells}, the copper is machined
either with a poly- or a mono-crystal diamond tool. Chemical
cleaning and etching is performed after the final machining step
to remove machining-related damaged material from the surface. An
iris and a half cup side forms a half cell,
Fig.\ref{figRDDScells}. After cleaning, the two half cells are
diffusion bonded together in a dry hydrogen furnace to form a full
accelerator cell. Several braze cycles are required for RF
couplers and water lines and a wet/dry hydrogen cleanup furnace
cycle follows the brazing steps. The structure is then vacuum
fired at 650$^{o}$C for two weeks to outgas dissolved dissolved
gases, primarily hydrogen, which can be available for breakdown
plasma formation. Vacuum-firing also reduces contaminant
concentration on the structure surface. Desorption of surface/bulk
gas contributes to the RF breakdown mechanism \cite{llaurent}.

Following electrical field mapping of the completed structure and
installation in the NLC test accelerator, the structure is
vacuum-baked, in-situ, for approximately one week prior to RF
processing. RF processing is required to ``season'' the structure
-- remove large particle contaminants, desorb gas from the surface
and vaporize any asperities by field emission. RF processing takes
place as a series of increasing pulse lengths and energies until
breakdown events at the desired operational gradient and pulse
duration cease. The objective of processing is to apply sufficient
power to physically alter potential field emitters without causing
damage, manifested as a phase shift, to the structure surface. A
typical RF processing procedure is as follows: Initial processing
takes place with 50ns pulses and the gradient is increased to 70
-- 75 MV/m. Once the structure operates without frequent breakdown
at that pulse width, the pulses are increased in length -- 100,
170, 240, and finally 400ns. After increasing the pulse length the
gradient is reset to a lower value and increased until consistent
RF performance is achieved at 75MV/m. Table~1 shows the variation
in RF Breakdown performance, for a steady operation, among several
structures at a lower gradient than the maximum gradient achieved.
RF processing times for similar structures have varied as much as
an order of magnitude with the range 100 to 1000~hours. Structures
that take less time to process tend to have better final
performance (fewer RF breakdowns.) This suggests two
possibilities: (1) Structures that required less RF processing had
fewer mechanical defects or surface/bulk contaminations
responsible for field emission or, (2) structures that require
longer processing times (due to outgassing or contamination)
undergo material changes that can subsequently produce field
emitters during operation.

\section{RF Breakdown Measurements and Structure Autopsy}

During catastrophic RF breakdown events, RF transmitted energy is
negligible \cite {dolgashev:linac2002}. Energy is partially
reflected and partially absorbed by the material surface. Hence,
breakdown events can be characterized by changes in the reflected
RF power. Approximate breakdown position in the structure is
calculated by the measurement of the the phase and timing of the
reflected signal. A complementary acoustic detection method, which
takes advantage of the conversion of the breakdown energy into
phonons, is also used to localize the breakdown events
\cite{Nelson:PAC03}.

RF-processed sections were cleanly autopsied (sectioned without
adding chemical contamination or physical debris). Scanning
electron microscopy (SEM) was used to obtain topographical images
of the surface and electron dispersive spectroscopy (EDS) was used
to determine the chemical composition of the surface and any
particle(s) that may be on or embedded into the surface.

Particle analysis was performed using Zep$^{TM}$ automated
searching software from ASPEX$_{LLC}$. Four strips per cell side,
each oriented azimuthally 90$^\circ$ apart and having $\sim$1.4~mm
width by $\sim$4.3~mm radial length from the edge of the iris,
were searched. Search conditions were set for feature size and
chemical composition. Any feature, which met the search criteria,
was detected by backscattered electrons during an initial survey
of the surface, then automatically magnified and imaged at 500x
using secondary electrons. Each imaged particle was also analyzed
by EDS.

\section{RF Breakdown in Travelling Wave Structures: Observations}

The site for enhanced field emission is some type of surface
feature, either intrinsic (grain boundaries, cracks, inclusions,
slip lines) or extrinsic (contamination particles, oxides, organic
residues). Contaminants, inclusions, and residues can be metal or
dielectric (field emission enhancement is known to occur, via
separate mechanisms, in both types of material.)

\begin{figure}[tbp]
\begin{minipage}[t]{.5\linewidth}
\centering
\includegraphics[width=0.9\textwidth,clip=]{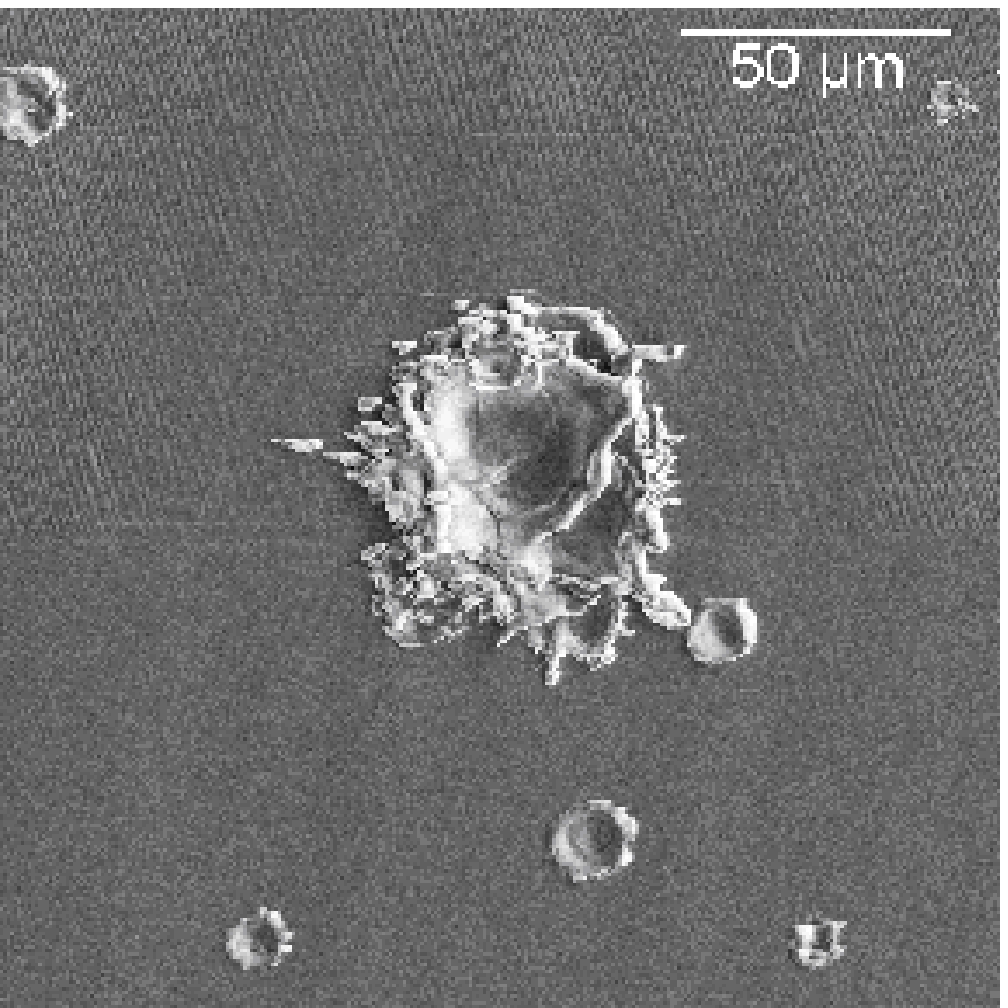}
\end{minipage}%
\begin{minipage}[t]{.5\linewidth}
\centering
\includegraphics[width=0.9\textwidth,clip=]{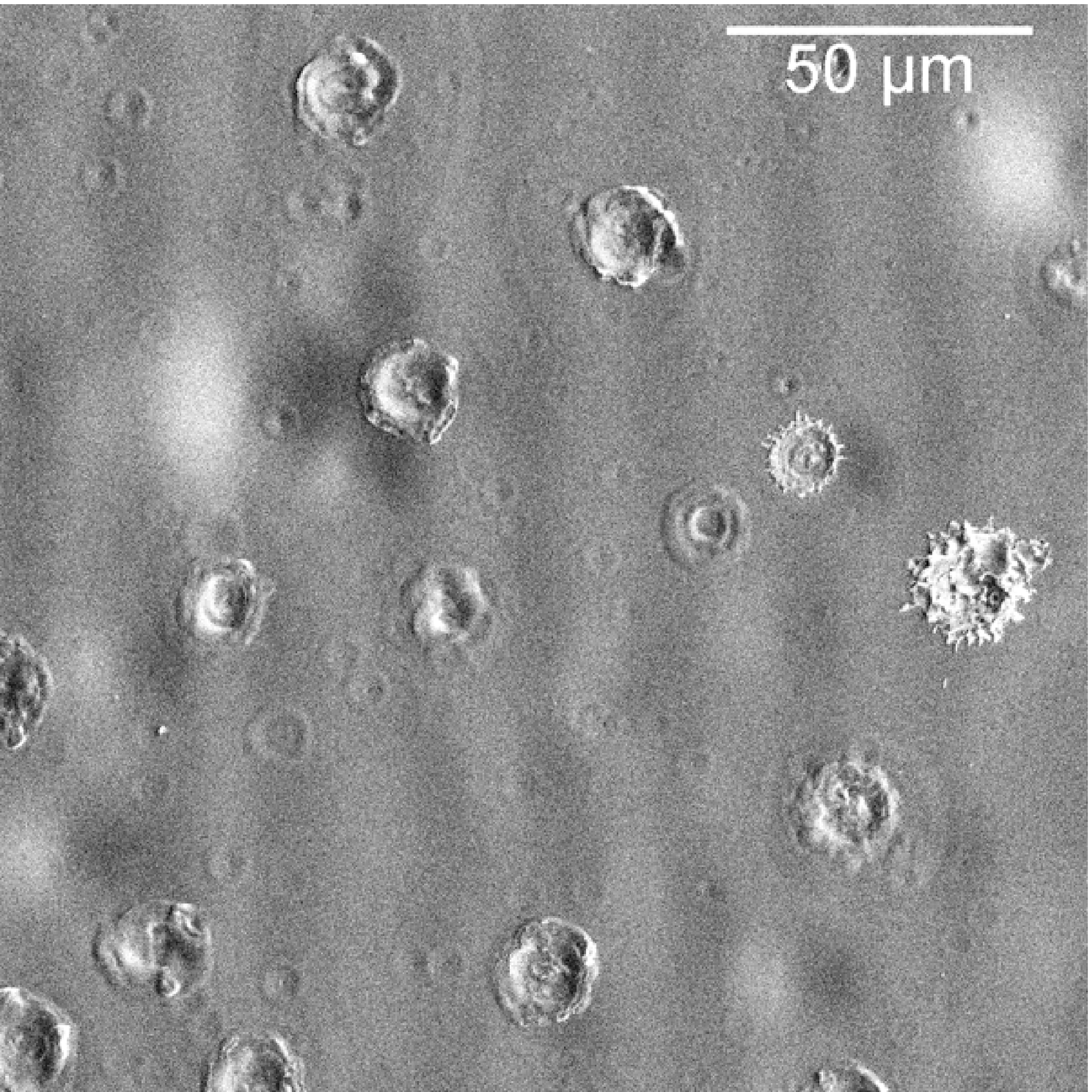}
\end{minipage}
\caption{(a) sharp-edged craters (left);  (b) smoothed-edged craters (right)}
\label{figsharpsmooth}
\end{figure}

The defect dissipates some of the RF energy. Melting at the
breakdown site creates craters that, due to their irregular
geometry, can act as further sites for enhanced field emission and
breakdown. The observed physical morphology of the craters falls
into two distinct groups: irregular edges and softened edges.
Fig.\ref{figsharpsmooth}(a) shows a crater with irregular edges;
in Fig.\ref{figsharpsmooth}(b) a second type of crater is observed
where there are few sharp edges and repeated melting appears to
have taken place. The softened-edge craters often appear in close
proximity with one another; irregularly edged craters are
frequently spaced further apart. This suggests that the
softened-edged craters may have started out irregularly edged, but
have changed morphology due to repeated melting or plasma-ion
bombardment.

In general the density of craters increases along the cell radius
toward the center, where electric fields (\textbf{E}) are highest,
Fig.\ref{figcraterdensity}. It is important to note that not every
grain near the aperture exhibited craters and that the crater
density varied substantially from grain to grain.

\begin{figure}[tbph]
\begin{center}
\includegraphics[width=0.8\textwidth,clip=]{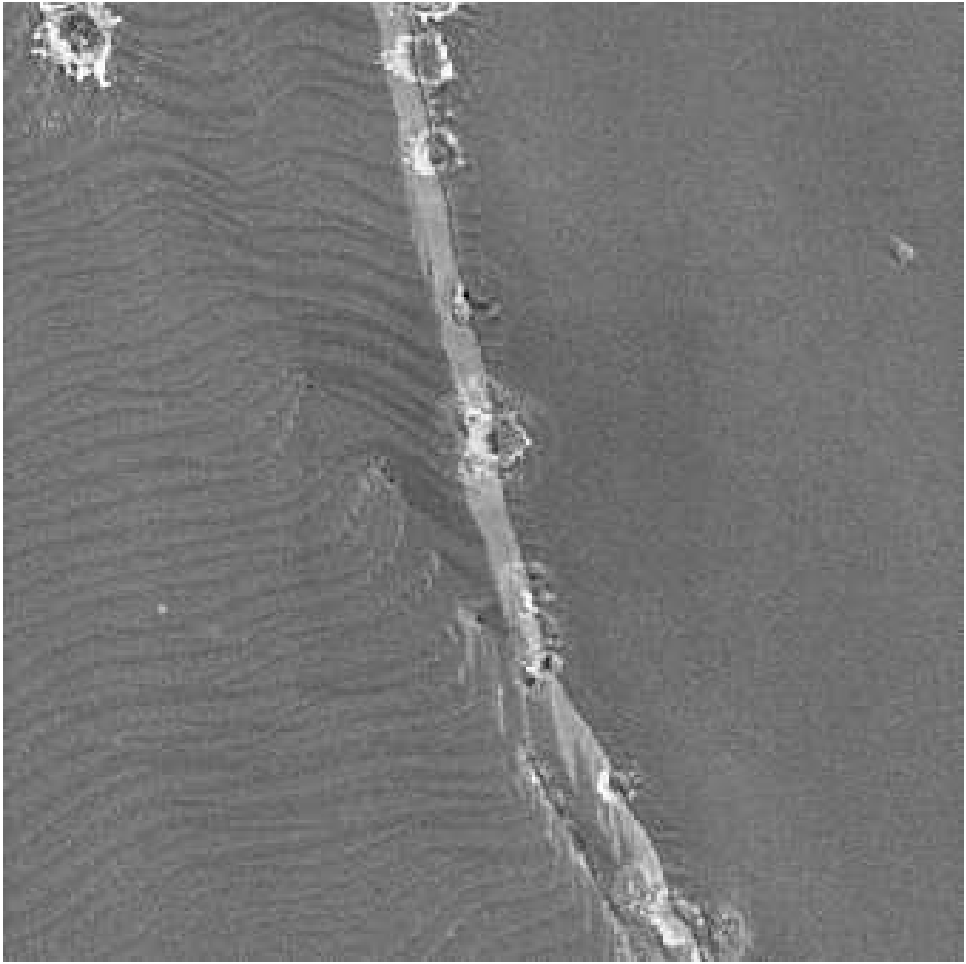}
\end{center}
\caption{Breakdown on Grain Boundary}
\label{figgrainboundary}
\end{figure}

Craters were also seen at grain boundaries and at inclusions,
Fig.\ref{figgrainboundary}. Grain boundaries and inclusions are
logical field emitters considering a Fowler-Nordheim model,
equation.\ref{FowlerNordheim}. Both of these defect types have
irregular geometries that increase the value of $\beta$
(field-enhancement factor); simultaneously these defects indicate
an area of high stress within the material, which could decrease
the work function $\Phi$, increasing the field emission:

\begin{equation}
J_{\left[ {A.m^{ - 2}} \right]} = \frac{e^3\,\left( {\beta \;E}
\right)^2}{8\;h\;\pi \;\Phi \;t^2(y)}\;\exp \left[ { -
\frac{4\;\sqrt {2\;m} \;\Phi ^{3 / 2}}{3\;\hbar \;e\;\left( {\beta
\;E} \right)}\;\nu (y)} \right]
 \label{FowlerNordheim}
\end{equation}

Changes in work function could explain the occurrence of RF
breakdown without the characteristic sharp edges and/or tall
asperities typically associated with high $\beta $. Preliminary
measurements of the local energy barrier height of the surface of
(111) OFE Copper via UHV-STM by Mizuno \cite{Mizuno:03} reveal a
value of 4.1eV, 18{\%} less than macroscopic work function value
of 5.0eV.

Of particular interest is the role of copper surface plasticity in
RF breakdown. Local heating, as well as the presence of high
magnetic and electric fields, are known to create mass movement in
copper \cite{Vorob:1989}. Inclusions emerging at grain boundaries
and microprotrusions (slip lines), created by local stresses due
to mass movement, can act as field emitters. So, there can be a
sufficiently high field or local heating that could induce mass
movement in copper, thus resulting in RF breakdown.
Fig.\ref{figgrainboundary} does present mass movement inside a
grain, and a crater with no traces of contaminant. However, we
have not correlated systematic mass movement in a grain with
breakdown events.

Mechanical deformation is known to change the local work function
in copper \cite{chow:1984}. This change in work function may
enhance Fowler-Nordheim field emission without increasing $\beta
$.

\section{T-Series Structure Autopsies: Input Coupler Design and Anomalous
Chemistry}

The first autopsies and SEM/EDS analysis took place on two test
structures, T53VG3R and T53VG3RA. The structures were nearly
identically manufactured; their difference being the final
chemical cleaning step. Prior to assembly, parts of the
R-structure were etched for 60 seconds; the RA-structure parts
were etched for 30 seconds. This corresponds to the removal of
3~microns and 1.5~microns of surface material, respectively
\cite{Kirby:LCC81}. The etch step removes material damaged by
machining, to some extent. The RA breakdown rate was twice that of
the R structure in the interior cells and 7 times greater when
including the breakdown attributed to the input coupler, Table~1.

Both structures exhibited high breakdown rates in their very first
cells due to breakdown events in the RF input coupler
\cite{lep_otpsy:Linac02}. The input coupler design had sharp
corners that amplified local heating and has since been
redesigned. It is speculated that breakdown events in the input
coupler create a shower of particles that travel into the
structure and created more breakdown events in subsequent cells.
Breakdown rates in the center cells of the structures were
significantly lower.

SEM autopsies of input, center, and output cells revealed numerous
craters. Craters were located at grain boundaries and in the
interior part of the grain. Inclusions were also observed. The
inclusions were metal-sulfide particles, particularly manganese
sulfide (MnS). The high number of MnS particles found in both
T53VG3R and RA are anomalous; the most commonly found particles
are copper, furnace insulation (silicon, calcium), aluminum,
carbon, and stainless steel (iron, chromium). MnS signal was
observed most frequently in cells close to the RF input coupler.
Normalized particle data is listed in Table~2 and Table~3. The
total area scanned per cell was $\sim$25~mm$^{2}$. The average
particle size in the interior cells was less than 1~micron.
Particles greater than 5~microns were seen most frequently in the
cell closest to the input coupler.

While greater particle counts were associated with the breakdown
events at the input coupler, there was little correlation between
particle density and breakdown events in a particular cell despite
the occurrence of MnS clusters in some cells. No correlation was
observed in autopsies of other systems.

\section{H-Series Structure Autopsies: "Hot" Cells due to Large Particles}

\begin{figure}[tbph]
\begin{center}
\includegraphics[width=0.8\textwidth,clip=]{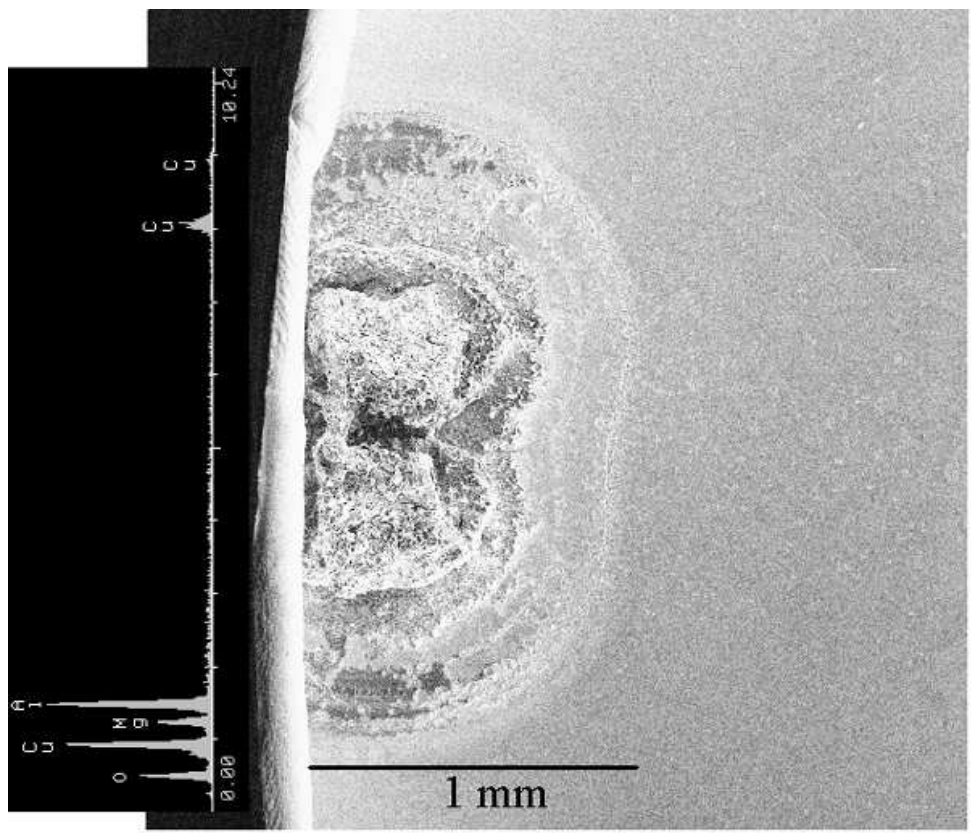}
\end{center}
\caption{Aluminum particle (melted due to RF breakdown?) found on cell 13 of H90VG5}
\label{figAlsliver}
\end{figure}

\begin{figure}[tbph]
\begin{center}
\includegraphics[width=0.8\textwidth,clip=]{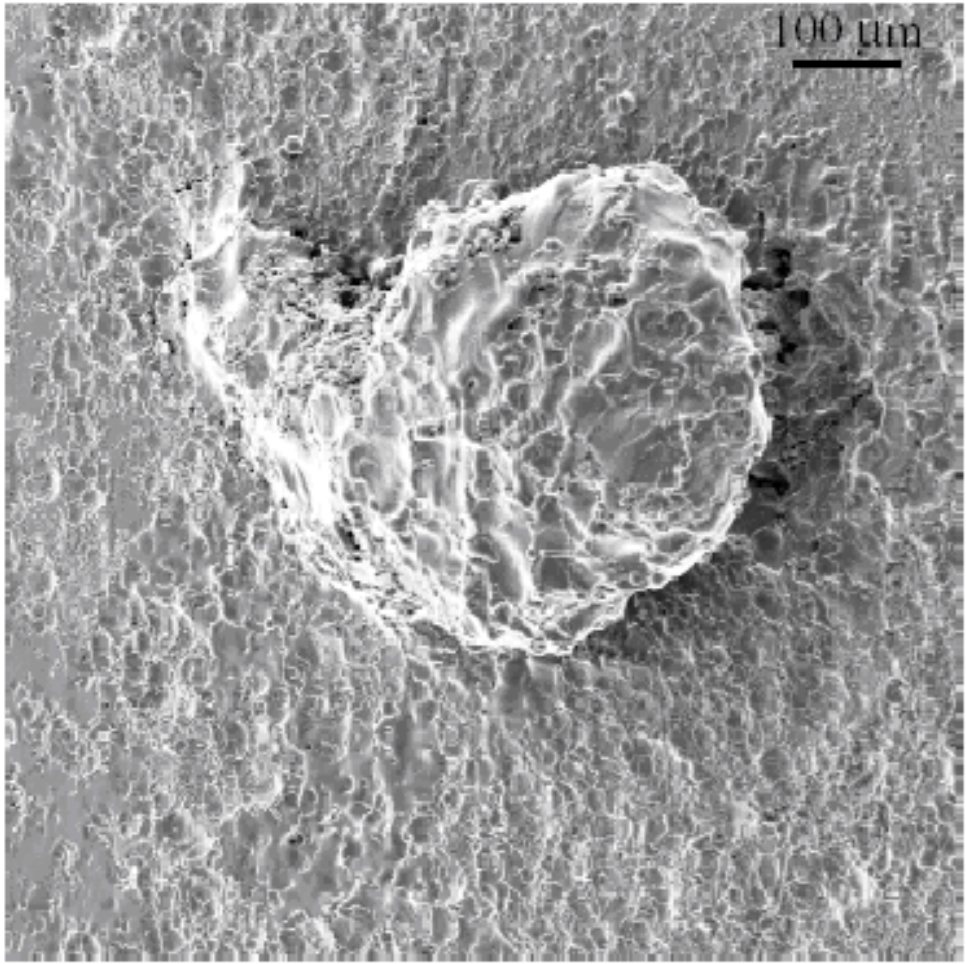}
\end{center}
\caption{Stainless steel particle embedded in wall of H90VG3N}
\label{figSSparticle}
\end{figure}

\begin{figure}[tbph]
\begin{center}
\includegraphics[width=0.8\textwidth,clip=]{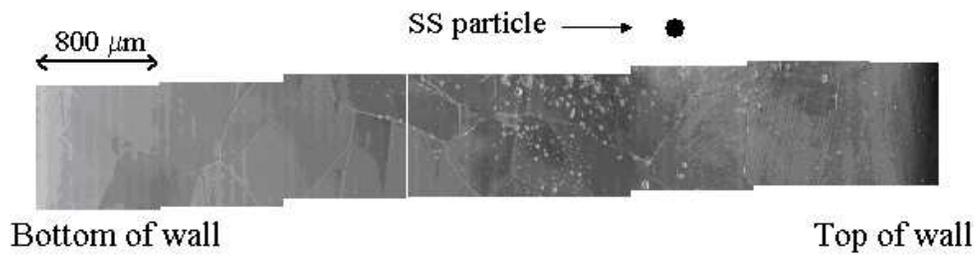}
\end{center}
\caption{High density of RF breakdown events near stainless steel mushroom}
\label{figcell35strip}
\end{figure}

Two processed structures, H90VG5 and H90VG3N, did exhibit high RF
breakdown in a localized region within the structure. At the
bottom of the cell 13, between iris 12 and iris 13, of H90VG5, an
aluminum sliver (mm length) was found, Fig.\ref{figAlsliver}. In
cell 35-36 of H90VG3N a large ($>$ 100~microns) stainless steel
(S/S) particle, Fig.\ref{figSSparticle} was found embedded in the
sidewall of the cell. A high density of craters was found near the
S/S particle, Fig.\ref{figcell35strip}. The aluminium sliver was
likely introduced to the structure during assembly in the NLTCA.
The process by which the S/S particle was introduced inside the
structure is not so clear because the particle is now actually
embedded into the copper.

Using RF design simulation software the intensity of the magnetic
or electric field inside a structure or a cell can be determined.
At the location of the cup side the magnetic field (\textbf{H}) of
the RF wave is maximal 0.3MA/m, on the iris it can be up to 5
times less . The surface electric field on the iris is 150MV/m and
is 7MV/m at the location of the S/S particle. The pattern of the
melting at the base of this particle is consistent with a high
current induced by \textbf{H} \cite{Dolgashev:private}.

H90VG3N was fitted with a new type of input coupler having rounded
edges; breakdown rates were high in cells nearest to the front of
the structure. Analysis of those cells shows particle densities
slightly lower than those in cells with significantly fewer
breakdown events, Table 4. Following the design change, the
highest number of breakdowns is concentrated at the entrance of
the first accelerating cells.

\section{Conclusion}

The field emission from particles and defects on accelerator RF
component surfaces can initiate irreversible breakdown damage in
high surface electric fields. Past electron microscopical analysis
of breakdowns at copper grain boundaries demonstrated the
importance of a sub-surface gas source for the breakdown plasma
creation, showing that vacuum bakeout of the finished structure is
useful in reducing this contributor to the breakdown process.

We focused in this work on the autopsy and SEM examination of
three fully RF-tested NLC structures, in order to determine
whether the large number of breakdowns seen in RF structures is
directly correlated with the particle load. The structure
autopsies show that near clean-room technique in fabricating
structures is essential to reducing gross particle loads that can
lead to breakdowns. Further, careful attention must be paid to the
structure vacuum bake environment in order to prevent adding
non-native contamination (for example, the evaporation of
manganese from the S/S vacuum components) to the last steps of
structure preparation. The autopsies, however, show that most RF
breakdowns have no visible particle residue or defect in the
resulting craters and that most particles do not, in fact,
breakdown. Therefore, most breakdown events must have some
not-yet-identified cause (presumably sub-surface) for the onset of
field emission. Studies of correlation between mass movement
inside a grain, hardness of grains with breakdown position, and
\textbf{(E,H)} field are ongoing.

\section{Acknowledgments}

We would like to thank C.~Adolphsen and V.~Dolgashev for useful
discussion, and G. Collet for developing the clean autopsy process
for structures.

\newpage


\clearpage

\clearpage

\clearpage

\clearpage

\clearpage

\clearpage

\clearpage

\end{document}